\pdfoutput=1 

\def\ifhtx{\iffalse}    
\ifhtx
\else
\documentclass[12pt,notitlepage,onecolumn]{ivoa}
\fi

\ifx\pdftexversion\undefined
  \usepackage[dvips]{graphicx}
  \DeclareGraphicsExtensions{.eps,.ps}
  \usepackage[ps2pdf]{hyperref}

\else
  \usepackage[pdftex]{graphicx} 
  \usepackage{epstopdf}         
  \makeatletter
  \g@addto@macro\Gin@extensions{,.ps}
  \@namedef{Gin@rule@.ps}#1{{pdf}{.pdf}{`ps2pdf #1}}
  \makeatother
 \usepackage[pdftex,bookmarks=true,bookmarksnumbered=true,hypertexnames=false,breaklinks=true,%
  colorlinks,linkcolor=blue,urlcolor=blue]{hyperref}
\fi
\usepackage[final]{pdfpages}
\usepackage{tabulary}
\title{Data Model for Astronomical DataSet Characterisation}
\date{2008 March 25}

\ivoatype{IVOA Recommendation}

\version{1.13}
\author{IVOA Data Model Working Group}
\editor{Mireille Louys, Anita Richards, Fran�\c{c}ois Bonnarel,
Alberto Micol, Igor Chilingarian, Jonathan McDowell\\}
\urlthisversion{\url{
http://www.ivoa.net/Documents/REC/DM/CharacterisationDM-20080325.pdf}}
\urllastversion{\url{http://www.ivoa.net/Documents/latest/CharacterisationDM.html}}
\previousversion{1.12}

\newcommand{\Sensitiv}{Sensitivity}

\newcommand{\blue}{\textcolor{blue}}
\newcommand{\violet}{\textcolor[rgb]{0.50,0.00,0.50}}

\begin{document}

\maketitle 

\section*{Abstract}
This document defines the high level metadata necessary to describe
the physical parameter space of observed or simulated astronomical
data sets, such as 2D-images, data cubes, X-ray event lists, IFU data,
etc.. The Characterisation data model is an abstraction which can be
used to derive a structured description of any relevant data and thus
to facilitate its discovery and scientific interpretation. The model
aims at facilitating the manipulation of heterogeneous data in any VO
framework or portal.

A VO Characterisation instance can include descriptions of the data
axes, the range of coordinates covered by the data, and details of
the data sampling and resolution on each axis. These descriptions
should be in terms of physical variables, independent of
instrumental signatures as far as possible.

Implementations of this model has been described in the IVOA Note available at:\\
\violet{\footnotesize{\url{http://www.ivoa.net/Documents/latest/ImplementationCharacterisation.html}}}

Utypes derived from this version of the UML model are listed and commented in the following IVOA Note:\\ 
\violet{\footnotesize{\url{http://www.ivoa.net/Documents/latest/UtypeListCharacterisationDM.html}}}

An XML schema has been build up from the UML model and is available at:\\ 
\violet{\footnotesize{\url{http://www.ivoa.net/xml/Characterisation/Characterisation-v1.11.xsd}}}
 
\section{Status of this document}
 This document has been produced by the Data Model Working Group.
It has been reviewed by IVOA Members and other interested parties, and has been endorsed by the IVOA Executive Committee as an IVOA Recommendation. It is a stable document and may be used as reference material or cited as a normative reference from another document. IVOA's role in making the Recommendation is to draw attention to the specification and to promote its widespread deployment. This enhances the functionality and interoperability inside the Astronomical Community.

\section*{Acknowledgements}
Members of the IVOA Data Model Working Group, including
representatives of the US NVO, Astrogrid, and the Euro-VO have
contributed to the present draft.  The editors particularly
acknowledge the input from P.\ Didelon, G.\ Lemson, A.\ Rots, L.\
Shaw, B.\ Thomas and D.\ Tody.\  F.\ Bonnarel and M.\ Louys
acknowledge support from the French ACI-GRID project, {\em IDHA} and
the EU-funded {\em VOTech} project.

\clearpage

\tableofcontents

\clearpage

\section{Introduction}
This document defines an abstract data model called ``Data Set
Characterisation'' (hereafter simply ``Characterisation''). In this
Introduction we present requirements and place the model in the
broader context of VO data models. In Section~\ref{sec:explore} we
introduce the concepts (illustrated with some examples) and
discuss their interactions. In Section~\ref{sec:model} we present
a formal UML class model using the concepts defined earlier. XML and
VOTABLE serializations are presented in Section~\ref{sec:serial} and
the Appendices give further examples.

\subsection {The purpose of the Characterisation model}
\label{sec:usecase}
Characterisation is intended to define and organize all the metadata
necessary to describe how a dataset occupies multidimensional space,
quantitatively and, where relevant, qualitatively. The model focuses
on the axes used to delineate this space,
including but not limited to {\it Spatial} (2D), {\it Spectral} and
{\it Temporal} axes, as well as an axis for the {\it Observable}
(e.g. flux, number of photons, etc.), or any other physical axes. It
should contain, but is not limited to, all relevant metadata generally
conveyed by FITS keywords.

Characterisation is applicable to observed or simulated
data\footnote{Unless otherwise stated, we use the terms ``dataset'',
``observations'' etc.  to mean any applicable observed or simulated
data.}  but is not designed for catalogues such as lists of derived
properties or sources (see Section~\ref{sec:othermodels}).

The model is intended to describe:
\begin{itemize} \item A single observation;
\item A data collection;
\item The parameter space used by a tool or package accessed via
  the VO.
\end{itemize}

The model describes the available data, not its history. For instance,
spatial resolution expresses the level of smearing of the true sky
brightness distribution in a data set without differentiating between
contributions from different atmospheric, instrumental and software
processing effects (see Section~\ref{sec:othermodels}).

 Characterisation has to satisfy two sets of requirements:
\begin{itemize}
    \item[I] Data Discovery requirements:

This model prescribes elements for use in requests to databases and
services and thus forms a fundamental part of the standards for VO
requests.  The use of this model should enable a user\footnote{A
user is either a human or a software agent} to select relevant
observations from an archive efficiently.   The selection will be
based purely on the geometry of the observations, that is, how and
how accurately the multidimensional space is covered and sampled.

Discovery may only require a simplified overview (e.g. position,
waveband, average spatial resolution). Data providers may opt for
the inclusion of data where there is insufficient information to
respond to certain parts of a query. Eventually, it should be
possible for a client to generate a detailed multidimensional
footprint of an observation. For example:
\begin{itemize}
        \item What observations from a particular archive are likely
        to have covered a specific VO Event? (Spatial and Temporal
        Coverage)
        \item Which CCD frames in a mosaic actually cover the position
      of a particular galaxy?  (detailed Spatial Coverage)
        \item What observed spectra have a resolution comparable with
      a given simulated spectrum e.g. matching the Shannon
      criterion? (Sampling Precision).
    \end{itemize}

\item[II] Data Processing/Analysis requirements:

Characterisation should detail the variation of sensitivity on all
relevant axes  (e.g. variation of sampling or sensitivity across the
field of view, detailed bandpass function), in order to  provide
information to an analysis tool or for reprocessing.

Errors may be provided for any or all axes.

\end{itemize}

 Version 1 will fulfill all Data Discovery requirements, and allow
some simple automatic processing such as cross-correlation and data
set comparisons. Full implementation of Data Processing/Analysis
requirements will only become available in a future version of this
model.

\subsection{Scope of the document}
    This document defines metadata items and organisation patterns for characterizing data products and their properties in the VO. It identifies some major contexts in which
these patterns play a crucial role and it shows how metadata
descriptions can be constructed in these contexts, in a form that is
adjusted to the requirements for distribution and analysis of astronomical observations. 
However, the precise application of
these patterns in other contexts may be different from these and we
do therefore as yet not prescribe the precise syntax of
characterization metadata in all contexts. That is left to the
controlling documents.

\subsection{Links to other IVOA modeling efforts}
\label{sec:othermodels} 
\begin{figure}[ht]
  \includegraphics[width=\textwidth]{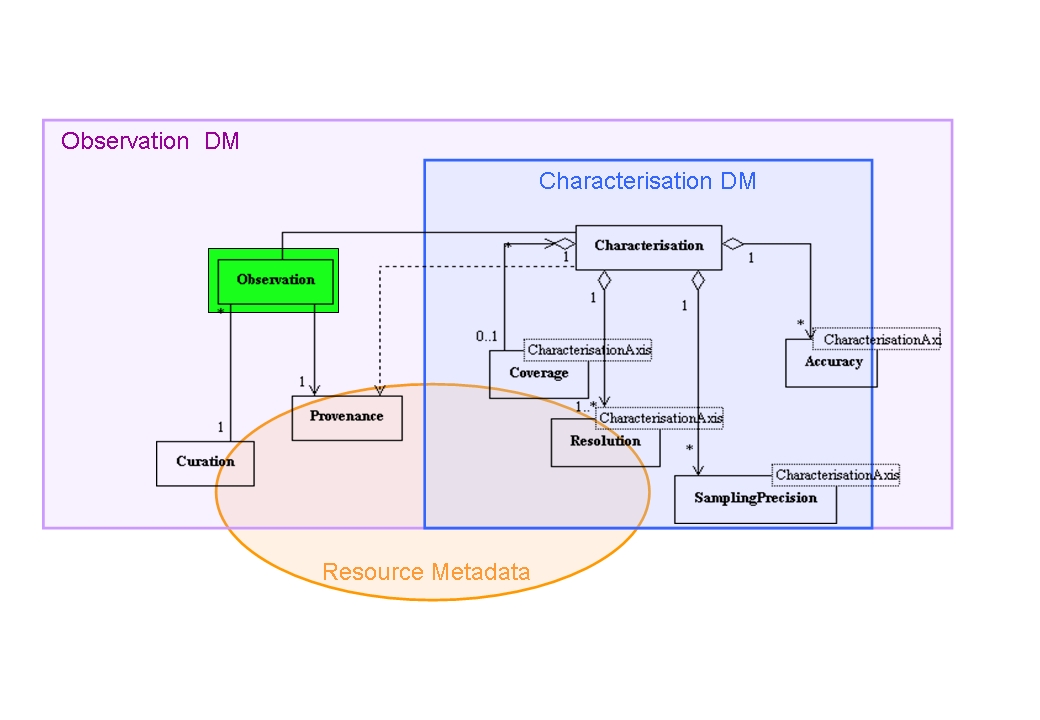}
  \caption{\emph{\small Interaction between the Observation and
  Characterisation data models: the Characterisation DM focuses on the
  physical information relative to an \textit{Observation} , a major class of the Observation Data Model.
  Characterisation DM has been identified as
  an important building block within the Observation Data Model, a large scope modeling effort
  to be completed in the future. Data management
  aspects such as VO identifier, data format, etc.. are handled in other metadata description
  like Resource Metadata or parts of the Observation model. }}
  \label{fig:generalObscharac}
\end{figure}

Characterisation arose out of the
``Observation Data Model'', a high level description of metadata
associated with observed data, described in an IVOA note available
at
\violet{\footnotesize{\url{http://www.ivoa.net/Documents/latest/DMObs.html}.}}
The connection is summarised in Fig.\ref{fig:generalObscharac}.  It
became obvious that there was an urgent need for a model to
characterise the physical properties of data, alongside Provenance,
DataCollection, Curation etc. (which provide instrumental,
sociological and other information). For example, Provenance will be
linked with Characterisation to provide the telescope location
(needed for some coordinate transformations), calibration history,
etc. Interactions are organised so that each Observation object of
the Observation data model will have a Characterisation object
encompassing the metadata relative to the physical axes along which
the measurements are spanned.

Characterisation complements and extends some of the metadata
adopted by the VO Registry
(\violet{\footnotesize{\url{http://www.ivoa.net/Documents/latest/RM.html}}}),
providing the finer level of detail needed to describe individual
datasets.
 Concurrently to the Characterisation DM, the Spectrum modeling effort appeared
(\violet{\footnotesize{\url{http://www.ivoa.net/Documents/latest/SpectrumDM.html}}}),
focusing on spectral data sets. It partly re-uses the
Characterisation metadata tree representation for data spanned along
the spectral axis. Data models for Catalogues and Sources are also
being developed.
Ideally, all these models must be mutually consistent and employ the
definitions supplied by the STC DM and/or by basic models describing
astronomical quantities as was planned within the Quantity DM.
However some overlap and duplication may happen to allow data and
service providers to use the parts they need without excessive
effort.

\section{Exploring the Characterisation concepts}
\label{sec:explore}
\subsection{Overview: a geometric approach}
\label{sec:concepts}
We introduce the physical axes used to define the N-dimensional space
occupied by any data set or required for interpretation.  When
considering a typical astronomical observation, we have identified
various Properties:

\begin{itemize}
\item {\it Coverage:} describes what direction the telescope was
  pointing in, at which wavelengths and when; and/or the region
  covered on each axis. This is described in increasing levels of
  detail (see Section~\ref{sec:cover}) by:
\begin{itemize}
\item {\em Location}
\item {\em Bounds}
\item {\em Support}
\item {\em Sensitivity}
\end{itemize}

If the data contain many small regions then the Bounds may be
qualified by a
\begin{itemize}
\item {\em Filling Factor}
\end{itemize}
(especially if the Support is not precisely defined).

\item {\it Sampling Precision:} describes the sampling intervals on each axis;
\item {\it Resolution:} describes the effective physical resolution
(e.g. PSF, LSF, etc.).
\end{itemize}
Each property can be related to one or more physical axes, described in more detail in Section~\ref{sec:props}. For each axis:
\begin{itemize}
\item {\it Accuracy:} describes the measurement precision, see
Section~\ref{sec:acc}.
\end{itemize}

\subsection{Examples of Characterisation}
\label{sec:examples} The tables below  illustrate how the spatial,
temporal and spectral domains and the observable quantity of some
typical data sets can be described, at various levels of complexity,
using the properties from Section \ref{sec:concepts}.
Table~\ref{tab:evt} shows some of the Characterisation metadata for
an X-ray event list. Additional examples are presented in Appendix C
: Table~\ref{tab:2Dim} for a 2-D image, Table~\ref{tab:1Dspec} for a
1D spectrum, Table~\ref{tab:3Dspec} for an IFU Dataset,
Table~\ref{tab:rad} for a radio interferometry image service and
Table~\ref{tab:sim} for simulated data.

In some of these examples, some concepts are interdependent,
discussed further in Section~\ref{sec:axesatt}.
All these concepts can be applied to any data set but some elements
may not have defined values, or the origin may be arbitrary, for example the spatial
location of a generic simulated galaxy cluster (Table~\ref{tab:sim}).

\begin{flushleft}
\begin{table}[htbp]
{\small
\begin{tabular}{|l|l|l|l|l|}
\hline
\space \textsc{\blue{Axes}}& \textsc{\blue{Spatial}} & \textsc{\blue{Temporal}} & \textsc{\blue{Spectral}}& \textsc{\blue{Observable}}\\
\textsc{\blue{Properties}} &           &                   &                  & \textsc{\blue{e.g. Flux}} \\
\hline \textbf{Coverage} \\ \hline
Location & Centra & Mid- Time & Central  & Average flux\\
         &position &           & energy &  \\
 & & & & \\
\hline
Bounds  & RA,Dec        & Start/stop   & Energy    & min:Probability \\
         &[min,max] or  &    time      & [min,max] & above  \\
         & Bounding box &              &           & background\\
        & [center, size]&              &           & max: Pileup\\
 & & & & \\
\hline
Support & FOV as accurate & Time      & Energy filter& \\
        & array of        & intervals & intervals & \\
        & polygons        &(array)    &(array)    & \\
 & & & & \\
\hline
Sensitivity&  Quantum          &  &ARF (effective  & Out-of-time  \\
            & efficiency(x,y); &  &area)as fn      & events \\
            & vignetting       &  & of energy      &(saturation); \\
            &                  &  &                &wings of PSF \\
& & & & \\
\hline Filling  & Good pixel &  Live time  & not &  \\
  factor        &fraction &  fraction      & used & \\
 & & & & \\
\hline
\textbf{Resolution} &  PSF (x,y)  & Time  & RMF (spectral  &SNR\\
          &  or its FWHM & resolution&  redist. matrix) & \\
 & & & & \\
\hline
\textbf{Sampling } &  Pixel scale & Frame   & PI bin       & ADU\\
\textbf{Precision} &  (x,y)       & time &   width & quantization  \\
 & & & & \\
\hline
\end{tabular}
}
\caption{\emph{Property versus Axis description of  metadata
    describing an \textbf{X-ray CCD Event List}. This also characterises the
    potential images and other products which can be derived. During exposure, the
    instrument moves with respect to the sky, so, for example, the
    sensitivity is a function of the support on the first three axes.}
} \label{tab:evt}
\end{table}
\end{flushleft}
\subsection{Structure and development strategy}
\label{sec:axis}
Characterisation provides a framework to present the metadata
necessary to specify a dataset in a standard format and to
make any interrelationships explicit. The description can be presented
from the perspective of the Properties or the Axes in a succession of
progressively more detailed description layers.  This will allow
evolution of the model in three independent directions: new properties
may be added as well as new axes, and if necessary new levels of
description may be considered without breaking the overall structure.

\subsection{The Axis point of view}
\subsubsection{Axes and their attributes}
\label{sec:axesatt}
 The physical dimensions of the data are described by axes such as:
{\sc Spatial}, {\sc Spectral}, {\sc Time}, {\sc Velocity}, {\sc
Visibility}, {\sc Polarisation}, {\sc Observable}.  We recommend that
data providers use these axes names but this is not compulsory
(e.g. FITS names can be used).  The data provider will be
\emph{required} to supply a UCD for each axis, as well as the
units.
This would help to disentangle ambiguous or unprecise metadata for
data retrieval or recognition by standard software. There is no
limit on the number of axes present and they may be dependent or
overlapping (e.g. one frequency axis and two velocity axes,
representing the velocities of two separate molecules with
transitions at similar frequencies).


Some axes may not even be explicit in the data, but are implicit,
present only as a header keyword or elsewhere. For example, a simple
2D sky image usually has celestial coordinate axes, but the time and
spectral axes may not be present in the main data array although the
observation was made using a finite integration time and wavelength
band (a single sample on each of the temporal and spectral axes).
These implicit axes may be represented in Coverage to provide their
location an/or bounds, or even, for purposes such as color
corrections, their sensitivity as a function of the coordinate within
the bounds.

\subsubsection{Axes flags}
\label{sec:axesflags} Axes flags (Section~\ref{sec:flagmodel}) are
used to indicate Boolean and other qualifying properties.  These
include whether the axis represents a dependent variable (e.g. the
Observable), the calibration status and whether the data are
undersampled.

\subsection{Accuracy}
 Accuracy characterises any uncertainties associated with each axis
(Section~\ref{sec:acc}) -- astrometric uncertainties are attached to
the Spatial axis, photometric to the Observable etc. Note that this is
a level of detail distinct from the assessment of the overall accuracy
of data provided by the Registry metadata.

\subsection{The Property point of view}
\label{sec:props}
The main properties needed for data description and retrieval are
categorized under Coverage, Resolution, and SamplingPrecision,
introduced in Section \ref{sec:concepts}.

The values of the properties characterising an Observation may be
derived from instrumental properties given in Provenance or from other
Characterisation features.  For example, high energy missions move the
telescope during the observation (Table~\ref{tab:evt}), leading to a
time-variable mapping from detector to celestial coordinates (the
`aspect solution'), giving a spatially variable effective exposure
time derived from the temporal bounds multiplied by the filling
factor, or the sum of all the support intervals weighted by
sensitivity, or derived from the sampling precision and period within
the bounds. The sensitivity across the spectral band may be a function
of spectral position (ARF).  Such dependencies should be restricted to
areas of significance to users, such as the Sensitivity class.  At
present, a single value, or the extrema, can be given for each
element; more complex formulae will be available in a future version
of Characterisation.

\subsubsection{Coverage}
\label{sec:cover}
\begin{figure}[htbp]
    \centering
    \includegraphics[width=\textwidth]{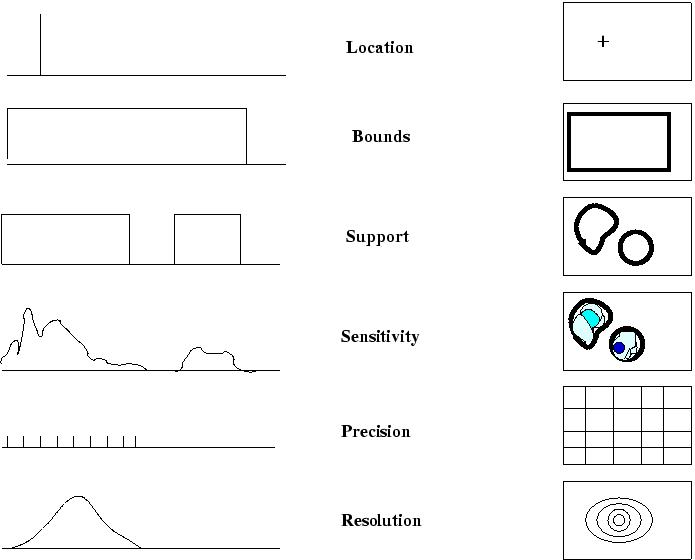}
    \caption{\emph{Illustration of the different levels of description.
    left: for a 1-dimensional signal, right: for a 2D signal. }}
\label{fig:charprinciple}
\end{figure}

 Coverage has several levels
of depth, providing a range of detail to meet the needs of
any user/developer, illustrated in  Fig.~\ref{fig:charprinciple}.
The simplest approximation to a spatial field of view presumes that a
sharp-edged region of the celestial sphere has 100\% sensitivity
inside and 0\% outside.  In reality the transition is fuzzy and the
region may be irregular and contain gaps. For example, some
applications only need to know what range of coordinate axes values
might contain data; others need to know the variation in (flux)
sensitivity as a function of position on an axis. Coverage provides
answers to these questions at different levels of precision, with the
idea that software implementations will be able to convert between the
levels.

 Coverage is described by four layers which give a hierarchical view of
increasing detail:
\begin{enumerate}
\item {\it Location:} The simplest Coverage element is the Location of
a point in N-dimensional parameter space, such as an image described
by a single value each of RA, Dec, wavelength and time. These are
fiducial values representative of the data. A precise definition
(mean, weighted median, etc.) is not required, but Location can serve
as a reference value or origin of coordinates in frames with no
absolute position (e.g. Table~\ref{tab:sim}).

\item {\it Bounds:} The next level of description is the
SensitivityBounds, i.e. a single range in each parameter providing the
lower and higher limits of an N-dimensional ``box''.  The scalar
intervals between the limits (the sizes and centres of each box-side)
should also be available if required.  The Bounds are guaranteed to
enclose all valid data but there may be excluded edge regions for
which there is no valid data, such as (on the wavelength axis) the
`red leak' end of a spectral filter.  These provisions satisfy the
intent of typical data discovery queries.

\item {\it Support:} Mathematically, the support of a function is the
subset of its domain where the function is non-zero. Here, Support
describes quantitatively the subsets of space, time, frequency and
other domains, onto which the observable is mapped, where there are
valid data (according to some specified quality criterion). Support
may include one or many ranges on each axis (e.g. Table~\ref{tab:3Dspec}).

\item {\it \Sensitiv:} Sensitivity, (unlike the previous `on/off'
properties), provides numerical values indicating the variation of the
response function on each of the axes, such as the relative
cell-to-cell sensitivity in the data. This includes filter
transmission curves, flat fields, sensitivity maps, etc.  The final
limits on Sensitivity are determined by the bounds of the Observable;
for example, the minimum and maximum given by a single count and by
the saturation level for some types of detector.
\end{enumerate}

The {\em Bounds} may also contain the
\begin{itemize}
\item {\it Filling Factor} sub-level, which gives the useful
fraction of Bounds on any axis. It may not be appropriate to detail
multiple small interruptions to data (for example detectors requiring
dead time between each sample) if it is conventional for analysis
systems to solve the problem using a statistical correction based on
the Filling Factor.  Very regular filling may also be described by
Sampling (see below).  Even if Support provides a complete
description, the Filling Factor may be used to rank the suitability of
data during discovery.
\end{itemize}

A method should be provided to derive the Filling Factor from the
Sampling Extent and Sample Precision (Section~\ref{sec:ressamp}) if
these are given, but if all three values are entered separately
there needs to be a means of checking for consistency.

\subsubsection{Resolution and Sampling Precision}
\label{sec:ressamp}
 Resolution is often a smoothly decaying (e.g. Gaussian) function but
the data product is subject to further discrete Sampling, e.g. CCD
pixels, Table~\ref{tab:2Dim}. Resolution may, however, be a top hat
function determined by the Sampling interval -- e.g. the temporal
resolution of an image made from a single integration.  We maintain a
distinction between the concepts to facilitate different requirements
in data processing, whether during data discovery services which allow
resampling or flexible resolution (Table~\ref{tab:rad}), or during
post-discovery processing (Table~\ref{tab:3Dspec}).

\begin{itemize}
\item {\it Resolution}
Resolution is usually the minimum independent interval of measurement
on any axis.
 Mathematically, if
the physical attributes (e.g. position, time, energy) of the incident
photons, or other observable,
are ${\bf x}$ (e.g.  $x_0$ = energy, $x_1$ = RA, $x_2$ = Dec, $x_3$ =
time, etc.), and the measured attributes are ${\bf y}$ (e.g. $y_1$ =
spectral channel, $y_2,y_3$ = pixel position, $y_4$ = time bin) then
given a flux of photons $S({\bf x})$ the detected number of photons is
\[ N(y_1,y_2,...) = N(\bf{y}) = \int S({\bf x})A({\bf x}) R({\bf
x},{\bf y}) dx \] where A is the probability that a photon is detected
at all (the quantum efficiency) and $R(x_1,x_2,...,y_1,y_2,...)$ is
the smearing of measured values (PSF, line spread function, etc.).

In the most detailed case, {\bf R(x, y)} may be a complicated
function, such as a PSF which varies as a
function of detector position and energy. The first level of
simplification is to specify a single function which applies to the
whole observation - e.g. a single PSF. This function may either be
provided as a parameterized predefined function (e.g Gaussian) or as
an array. The concept of Resolution Bounds provides the extreme values
of resolution (see Table~\ref{tab:rad})

The final level of simplification is to give a single
number characterising the resolution, such as the
the standard deviation of a Gaussian PSF.

\item{\it Sampling}

Sampling (or pixelization or precision or quantization) describes the
truncation of data values as part of the data acquisition or data
processing. If sampling is non-linear, simplification may be
necessary, by giving limiting values or a single `characteristic
sampling precision'. The Sampling Period gives the sample separation
and the Sample Extent shows the deviation from the pure ``Dirac comb''
case.  The Nyquist parameter -- the ratio between the resolution FWHM
and the Sampling period -- will also be provided by a method.  The
Sampling flags (Section~\ref{sec:flagmodel}) provide a simple guide as
to whether these properties are significant.
\end{itemize}

\subsection{Presentation of layered information}
\label{sec:layer} The layered structure allows tasks to retrieve
only the metadata which is actually required. The lower levels can
be very detailed, for example the variation in Sensitivity to the
Observable(s)  along the spatial, spectral and other axes, or the
variation of the resolution within the field of view.
 This could
take various forms:
\begin{itemize}
\item A simple value or range
\item An analytic function of other property values
\item A variance map for 2D data
\item A look-up table for the bandpass correction to 1D spectral data
\end{itemize}

The more complex properties may be provided using pointers to
ancillary data with the same types of axes and dimensions as the
observation itself, e.g. a weight map packaged with a 2D image; this
capability exists in the first version of this model.  The provision
of ``attribute formulae'' or attributes pointing to functional
descriptions, such as the aspect solution for an X-ray observation,
is left for the future development of Characterisation; a first step
may be to decompose a complex coupled description into non-coupled
expressions. Where it is possible to provide separate values for
interdependent elements (see also the end of
Section~\ref{sec:cover}), there must be a validation method to avoid
contradictions.

 A later version of the model will also allow links to other aspects
of the Observation model, external calibration and documentation.
Advanced VO tools could use such metadata to recalibrate data on
demand.  Characterization is used to describe potential as well as
static data products (e.g. Tables~\ref{tab:evt} and~\ref{tab:rad}).
It could therefore also provide pointers to Registry entries
indexing tools and services that could be launched on the fly for
extracting images etc. from event or visibility data or atlas
cut-outs.

\section{The Model}
\label{sec:model}
\subsection{The role and structure of the Model}
\label{sec:struc}
\begin{figure}[ht]
    \centering
        \includegraphics[width=\textwidth]{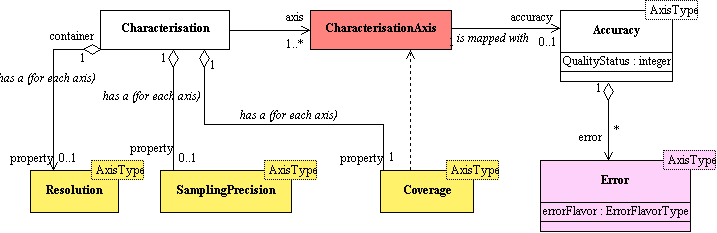}
        \vspace{0.3cm}
        \caption{\emph{ This UML class diagram emphasises the
        Property/Axis perspective. The Characterisation class is a
        container that gathers the properties for each axis. The axis
        is described by the CharacterisationAxis class.
        All relevant axes for one observation/dataset are linked to the Characterisation class.
        The AxisType template parameter for each Property allows to link properties to the corresponding Axis.
        The Accuracy class, linked to the CharacterisationAxis class, gathers different types of Error
        descriptions (systematic, statistical) as well as quality flags.}}
        \label{fig:axisframeAndProperties}
\end{figure}

\begin{figure}[htbp]
    \centering
        \includegraphics[width=\textwidth]{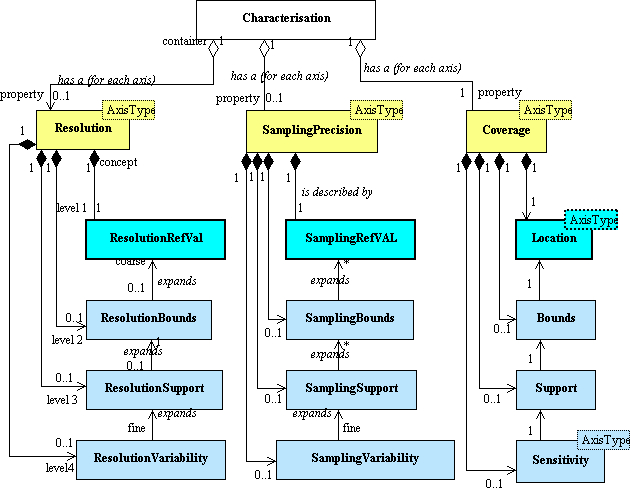}
        \caption{\emph{The layered structure of Characterisation:
        This diagram synthesises the Property/Axis/Layer approach.
        The concepts are represented in yellow. The
        coarse description is designed by the first level (blue boxes), while the
        pale blue ones represent the complementary metadata. The Bounds, Support and
        \Sensitiv~ classes are nested levels of detail to add knowledge
        about the Coverage of an Observation. Symmetrically, Resolution and
        Sampling may also have the 4-level structure of description.
        The  complete Characterisation for one observation is
        obtained by filling the tree for each relevant axis: spatial, spectral, temporal,
        etc.}}
        \label{fig:simplecharac}
\end{figure}

We use UML diagrams to describe the organisation of Characterisation
 metadata following the Properties/Axis/Levels perspective.  The model
 offers different views of the characterisation concepts.  Figure
 \ref{fig:axisframeAndProperties} shows the relationships between the
 main concepts. The AxisType box attached to each property
 class represents the axes along which the property (e.g. Resolution)
 is assessed; for example, there can be one Resolution class for each
 relevant axis.  Fig~\ref{fig:simplecharac} illustrates how the
 properties of the data are gathered under the Characterisation
 container class.  The Coverage class is shown with the four
 increasingly detailed properties introduced in
 Section~\ref{sec:props}; such a Characterisation tree is
 available for each type of axis.


\subsection{Axis description}
\label{sec:af} All the information related to an axis is gathered
within the CharacterisationAxis class. This can have common
``factorised'' attributes applicable to the property layers on that
axis (Section~\ref{sec:concepts}). It contains the name of the axis, units, UCD as well as a holder
for the STC coordinate frame (see Section~\ref{sec:QSTC}) which also
provides the base class for the observatory location (Observation --
Provenance model).

If a deep level (higher number, Section~\ref{sec:props}) object,
e.g. \Sensitiv, needs to have its own axis description, this can be
defined locally, overriding the factorised top level
CharacterisationAxis object. The redefinition can be partial, e.g. a
change of unit or a change of spatial orientation requiring a new
CoordSystem element.

\subsubsection{Flags and other qualifying information}
\label{sec:flagmodel} Other elements in the CharacterisationAxis
class include the number of bins present on this axis, and flags to
indicate the calibration status, independency and sampling
properties of the axis, as described in Section~\ref{sec:axesflags}

\paragraph{Independent or dependent status}
Axes may include both `independent' variables (which may have
associated errors) and the \emph{``Observable''} axis or axes which
represent phenomena measured along some other axes.  For instance,
in a 3D datacube of the sky, the Spatial axis is an independent axis
(flag TRUE), as is the (implicit) Spectral axis, but the Flux axis
is dependent (flag FALSE), and the velocity axis is dependent on the
frequency axis.

\paragraph{Calibration status}
The CharacterisationAxis object in the Characterisation model
provides a calibration status flag for each axis, so that a user can
insist on calibrated data only where necessary.  The
CalibrationStatus is given separately for each type of
characterisation axis and can be
\begin{itemize}
\item UNCALIBRATED: not in units which can be directly compared with
  other data (but often still useful, for example the presence of
  spectral lines at known wavelengths can give a redshift regardless
  of absolute flux densities).
\item CALIBRATED: in reliable physical units or other accepted units
  such as magnitudes.\footnote{In such cases the coarser levels of
  description should also be given in physical units and the need for
  a tool such as a look-up table of zeropoints etc. and conversion
  algorithms has been identified.}
\item RELATIVE: calibrated to within a constant
  (additive or multiplicative) factor which is not precisely known,
  such as arising from uncertainty in the flux density of a reference
  source.
\item NORMALIZED: dimensionless data, divided by another data set (or
  a local extremum).
\end{itemize}
The calibration process itself is described elsewhere in
the Observation Data Model (Section~\ref{sec:othermodels}).

\paragraph{Sampling status}
\begin{itemize}
\item Undersampling: TRUE if the sampling
precision period is coarse compared to the resolution and the
precision of a single data value is limited by the sampling; FALSE if
the sampling precision period is small compared to the resolution and
precision is limited by the resolution
\item Regular sampling: TRUE if the pixellation or binning is close to
  linear with respect to the axis world coordinate (so that an
  accurate position can be obtained by counting samples from a Bound);
  FALSE if this would introduce an error significant with respect to
  other uncertainties.
\item
The total number of samples along each axis may be given, normally
used for multiple regular sampling.
\end{itemize}

\subsubsection{Errors in Characterisation: the Accuracy class}
\label{sec:acc}
\begin{figure}[htbp]
    \centering
        \includegraphics[width=\textwidth]{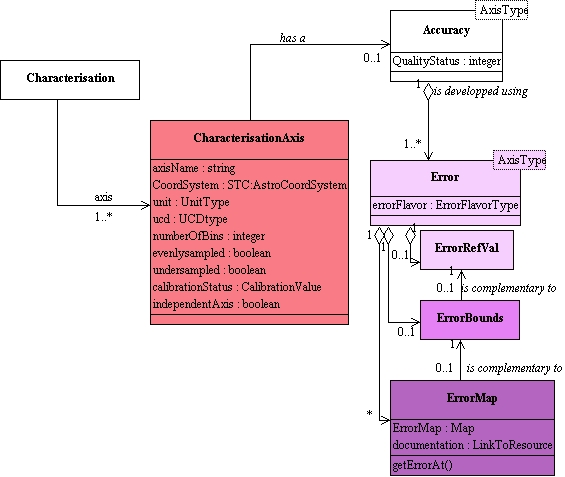}
        \vspace{0.3cm}
    \caption{\emph{This class diagram illustrates the CharacterisationAxis class
    and its relationship with the Accuracy  class, which encompasses
various types of errors such as systematic or
    statistical.  }}
    \label{fig:error}
\end{figure}

The values along Coordinate axes and measurements of Observables may
all suffer from systematic and statistical uncertainties.  Errors
may be in the units of the axis or may be represented by quality
flags. These Error classes are gathered in an Accuracy object
(linked to the CharacterisationAxis object, see
Fig.~\ref{fig:error}, and STC data model elements, see
Section~\ref{sec:QSTC})). Accuracy supports multiple levels of
description, analogous to Coverage.  The uncertainty in the position
or measurement on any axis can be described by a typical value, by
the bounds on a range of errors, and/or by very detailed error
values fo\textsl{}r each sampling element (e.g.
pixel).\footnote{Measurement errors are distinct from any
`fuzziness' in the values provided by the coarsest levels of
Characterisation, e.g. Location may be an arbitrary approximation
(Section~\ref{sec:cover}), but that kind of uncertainty is catered
for by going to deeper levels of Characterisation, and by the
concept of Region of Regard in the Registry Resource model.} A
pointer may be provided to error maps packaged with the data, as
described for the more detailed levels of Coverage.

\subsection{Navigation in the model: by axis or by properties?}
\label{sec:flex}
The structure of Characterisation is clearly hierarchical with the
characterisation class as the root element.  The model can be
serialised using two alternative sets of primary elements:
\begin{itemize}
\item {\em Properties}, with the corresponding classes for each axis
  attached; used, for example, to represent data where the axes
  values are interdependent (e.g. Table~\ref{tab:evt});
\item {\em Axes}, factorising each description into the multi-layer
  property levels; this provides more compact XML.
\end{itemize}
Either structure could be applied to the examples tabulated in
Section~\ref{sec:examples}.  This UML model could be used to build two
different XML schemas, giving access primarily by property or by axis.
Here, we present the ``Axis First'' serialisation only; the ``Property
First'' serialisation will be presented in the next version of this
model.

\subsection{Implementing the model using STC elements}
\label{sec:QSTC}
STC, the metadata scheme for Space-Time Coordinates (see \\
\violet{\footnotesize{\url{http://www.ivoa.net/Documents/latest/STC.html}}})
encompasses the description of most of the Characterisation axes
examples in Section~\ref{sec:axesatt} with the exception of
Observable. \Sensitiv~ is the only Property not present in STC.
However, the full STC structure cannot simply be reused, as it does
not have the flexibility needed to deliver the alternative schemata
for both multi-layered views presented in Sections~\ref{sec:struc}
and~\ref{sec:flex}. We do use STC intermediate level objects as
building blocks for the Characterisation model.

The STC:AstroCoordSystem object is needed as a reference to define the Coverage axes.
STC
substructures may be reused in the following way:

\begin{itemize}
\item {\em Location} implements STC:AstroCoords
\item {\em Bounds} encapsulates STC basic types, some STC:Interval elements and STC:Coords into a structure similar to STC:AstroCoordArea.
\item {\em Support} uses STC:AstroCoordArea
\item {\em Resolution} ResolutionRefval can be implemented via adhoc types using STC:CResolution elements
\item {\em SamplingPeriod and SampleExtent} encapsulate CPixSize elements from STC.
\end{itemize}

\begin{figure}[htbp]
    \centering
        \includegraphics[width=\textwidth]{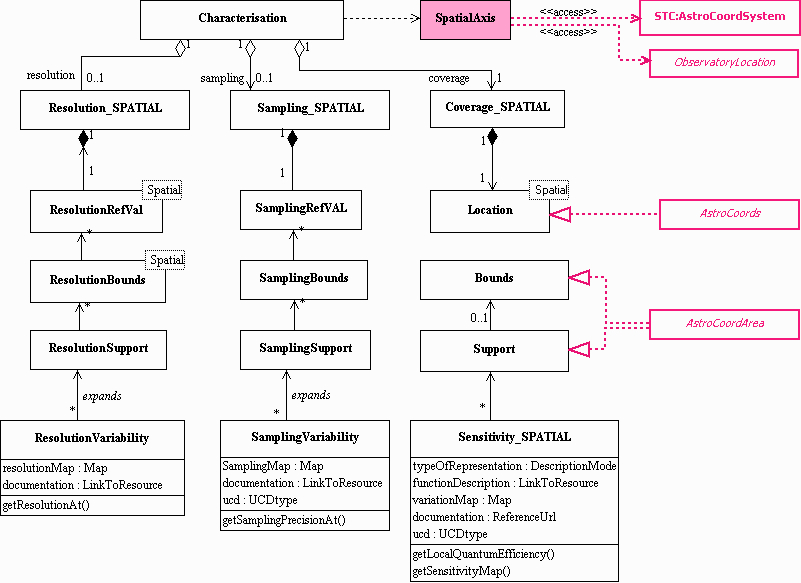}
        \vspace{0.3cm}
    \caption{\emph{Expressing the spatial properties as a subtree of Characterisation .
    Here is an example of how STC components (in pink italics) may be used to
implement the different levels
    of the Coverage description. The Location element uses a STC:AstroCoords. Bounds encapsulates STC
    basic structures like STC:Interval elements and STC:Coords in a structure similar to STC:AstroCoordArea.}}
\label{fig:spatialcharacsysobsloc}
\end{figure}

This is represented for the spatial axis using implementation links in
the UML diagram in Fig.\ref{fig:spatialcharacsysobsloc}.

In simple cases data handlers will probably reuse predefined
elements included from an external STC library.  For example,
CharacterisationAxis includes the STC elements for CoordSys and the
(possibly variable) space-time coordinates of the
ObservatoryLocation\footnote{This should, where necessary, be
consistent with the Provenance section of the Observation model
(Section~\ref{sec:othermodels}).} or of the origin of coordinates
(e.g. for barycenter-corrected data).

Many parameters (i.e. most numerical-valued elements at a finer
level than Location) are customarily expressed either as maximum and
minimum values or as a centre and scalar range (or both).  In some
cases an array of such values is needed, e.g. 2 dimensions on the
spatial axis in most but not all cases; upper and lower bounds to
(separately) the major and minor axes of Resolution in
Table~\ref{tab:rad}; higher dimensionality is possible such as the
inclusion of beam position angle in this Resolution example.

The Resolution and Pixel-Size concepts are represented in STC at a
deep level inside the Coordinates class (together with the
Name/Value/Error in the Coordinate object). This allows any coordinate
to be expressed to the appropriate degree of numerical precision.
Characterisation  needs to allow selection of metadata by
resolution, which therefore must be accessible at the upper level of
\textbf{description and is coded as a Property along one
CharacterisationAxis, as well as SamplingPrecision.}

Since the space, time and spectral axes are particularly important for
astronomy, we recommend that implementations include a method to
return an STC::AstroCoordSys object, which will only succeed if a
complete and consistent space-time-spectral description is present.
This may be nominal or arbitrary for some axes e.g. for simulated
data.

\section{XML Serialization}
\label{sec:serial}
\subsection{XML schema (Axis First)}
\subsubsection{Design of the schema}
Due to the hierarchical nature of the Model, the XML serialization
of Characterisation is based here on a single tree. The appropriate
elements are taken from STC as described in
Section~\ref{sec:QSTC}.  The root element called
``Characterisation'' is the aggregation of a set of
CharacterisationAxis elements\footnote{These elements are containers
gathering the result of the dynamic grouping of properties for a
given characterization axis} for each of the axes.  The
CharacterisationAxis element contains all axis information like an
obvious label (``spatial'', ``temporal''), coordinate system, units
, etc. Coverage implements different elements according to the four
levels of description detailed in Section~\ref{sec:cover}.
Lower levels of these properties along one particular
CharacterisationAxis may reuse the axis parameters defined into the
top-level objects for that axis or redefine their own axis
parameters(units, coordsystem, ...) locally, as described in Section~\ref{sec:af}.

A full XML serialisation is provided, as an XML schema, for simple
observations, at the following site:\\
\violet{\footnotesize{\url{http://www.ivoa.net/xml/Characterisation/v1.11}
}}\\
An XML instance document \footnotesize{MPFS-v1.11.xml}  describing an IFU dataset characterisation
is available at\\
\violet{\footnotesize{\url{http://www.ivoa.net/internal/IVOA/CharacterisationDataModel/}}.}

Most of the usual (spatial, time, spectral, velocity) coordinates
information re-uses the STC coordinates definitions and structures.
For simple values, the present version of the schema makes use of
STC fine grain structures like CPixsize or CError. Such terminal
leaves elements in the Characterisation XML tree could be adapted
and re-use simple existing structures (value, unit, semantic tag,
coding format) when available in a standard.

\subsubsection{Building blocks of the schemata}
In order to illustrate how the XML schemata is derived from the UML
Model, building blocks of the Schemata, corresponding to some main
classes of the UML diagram are shown here.

 The principle is to map
the main classes in XML elements, building up a hierarchy from the
most englobing concept down to  more specific ones. Aggregated
classes are easily translated as aggregated subelements. The
attributes of an UML class are also coded as sublevel elements.

\begin{figure}[htbp]
    \centering
       \includegraphics[width=0.75 \textwidth]{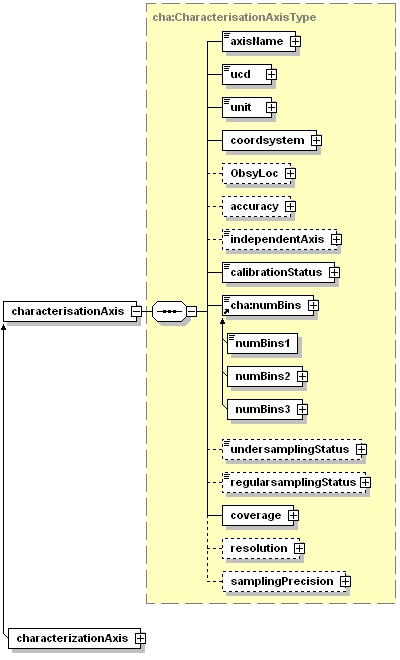}
\caption{\emph{\small The \textbf{CharacterisationAxis} element is built up
following the corresponding UML class with \emph{coordsystem} and
\emph{ObsyLoc} items reusing STC elements. The small arrow on
\textit{cha:numbins} represents a substitution group head element in
XML. This allows to plug various constructs of this element (e.g.
for 1D, 2D, 3D) that play the same role in the XML tree.}}
\label{fig:axisframeschema}
\end{figure}

\begin{figure}[htbp]
    \centering
       \includegraphics[width=0.82 \textwidth]{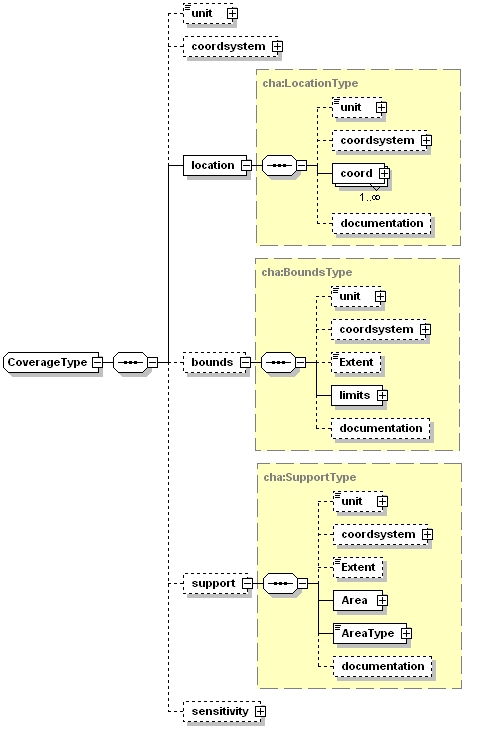}
\caption{\emph{\small The \textbf{coordsystem} and \textbf{unit} items can be
        factorised at the top of the Coverage structure, but may be
        redefined at each level when necessary. \textbf{Bounds} are expressed using a \textbf{limits}
        element which is developped on a general bounding box
        type: \textbf{CharCoordArea}.
        \textbf{AreaType} is a string describing the kind of region used: \textbf{Circle},
\textbf{Polygon} etc.}} \label{fig:coverageschema}
\end{figure}

The translation from UML to XML used in this serialisation applies
rules and elaborates specific techniques very similar to the work of
Carlson (\emph{Modeling XML applications with UML},
Addison-Wesley, 2001). The examples shown here are
`handmade' translations of the UML model.
Automated translation will be discussed in the next version of
Characterisation. The derivation of the XML from the UML model
is expressed in the graphical views of the XML schema in Figs.
\ref{fig:axisframeschema}, \ref{fig:coverageschema},
\ref{fig:resolschema}, \ref{fig:samplingschema} and
\ref{fig:accuracyschema}.

\begin{figure}[htbp]
    \centering
       \includegraphics[width=0.8\textwidth]{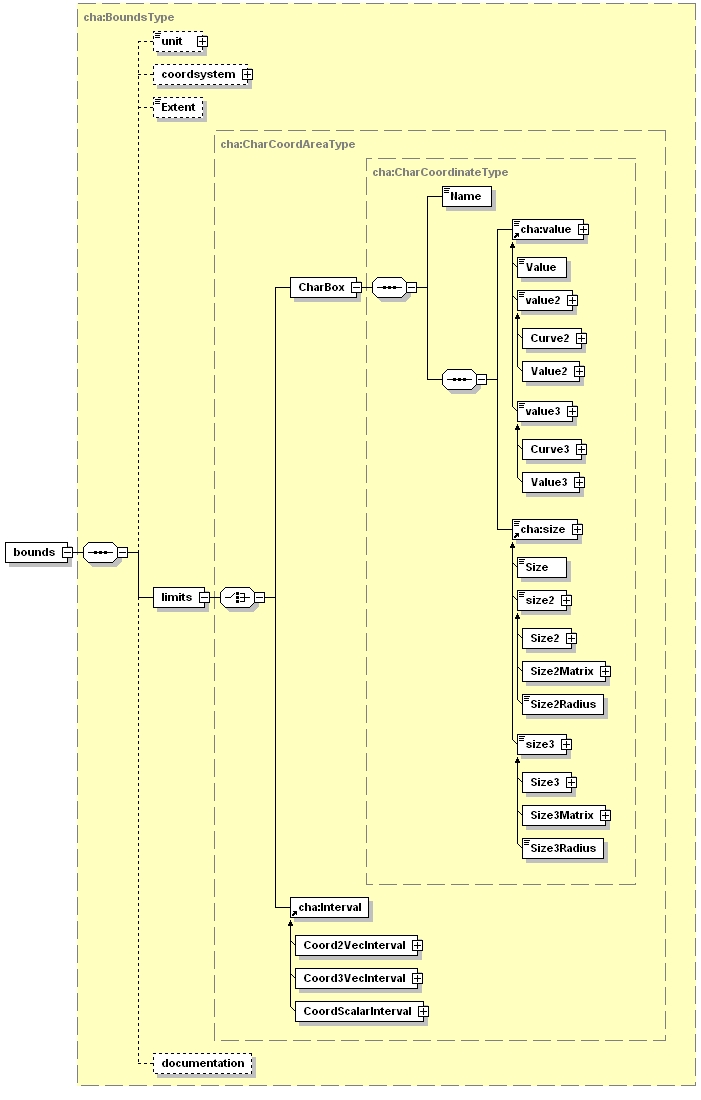}
\caption{\emph{\small Representing limits: The two expressions allowed for
a bounding box are expressed using either a STC:CoordInterval
embedded in a locally defined type cha:Interval or built on another
type: \textit{CharBox} representing a generic centered box in
N-dimensions.}} \label{fig:limitsschema}
\end{figure}
\begin{figure}[htbp]
    \centering
       \includegraphics[width=0.82\textwidth, height=0.8\textheight]{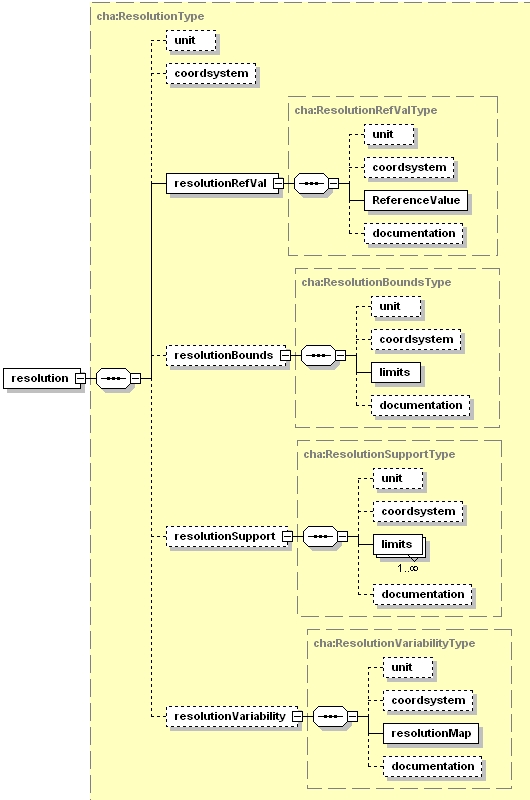}
\caption{\emph{\small This graphical view was generated with XMLSPY from
the resolution element of the schema. As designed in the UML class,
the resolution item may contain 4 possible subelements. The RefVal
element should be present but is not mandatory: some observations
may have unknown resolution.} \label{fig:resolschema}}
\end{figure}

\begin{figure}[htbp]
    \centering
       \includegraphics[width=\textwidth]{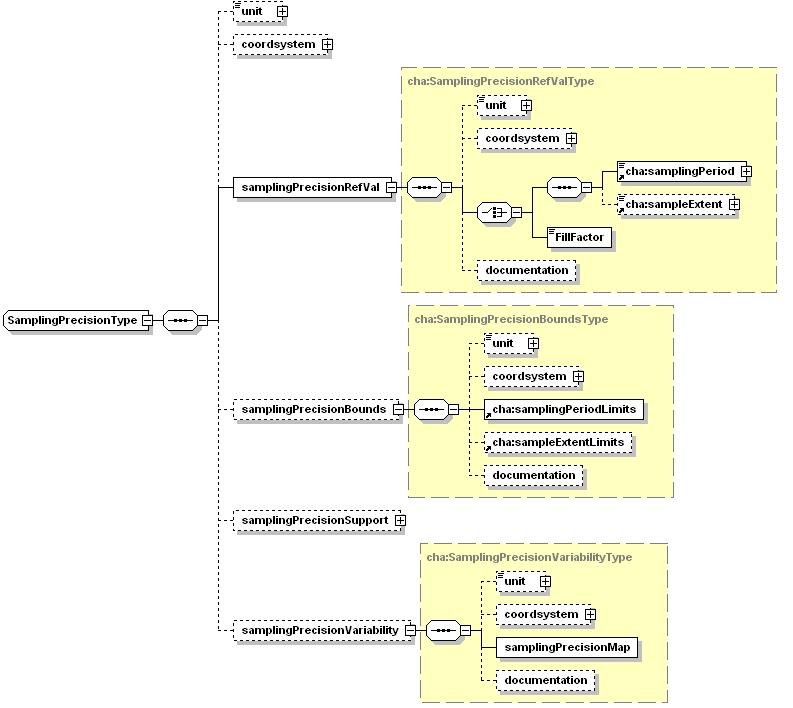}
\caption{\emph{The samplingPrecision item contains 4 possible
subelements. One among SamplingPrecisionRefVal and
SamplingPrecisionBounds should be present when possible but this is
not explicitly described by the XML syntax.}
\label{fig:samplingschema}}
\end{figure}

\begin{figure}[htb]
       \includegraphics[width=\textwidth]{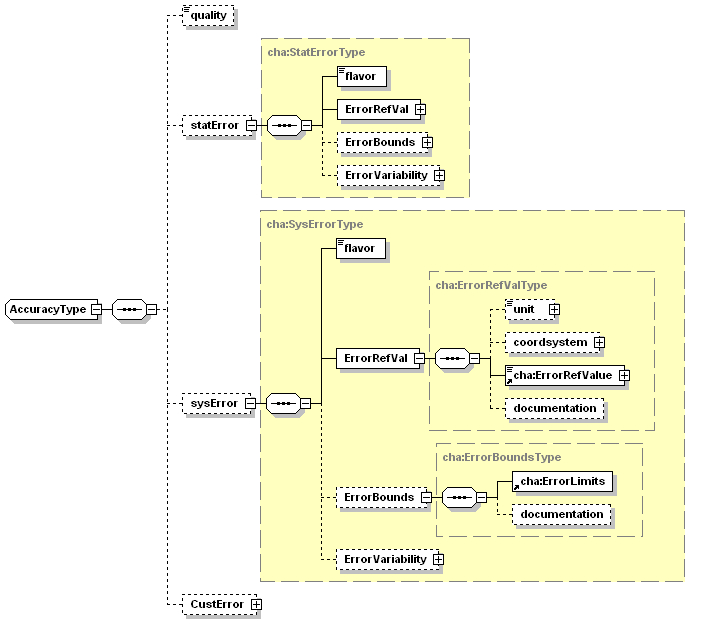}
       \vspace{0.3cm}
\caption{\emph{The accuracy element relies on Errors along the axes
and is built up on STC elements.} \label{fig:accuracyschema}}
\end{figure}
\clearpage

\subsection{Utypes generation: select one ordering strategy}
One application of such a model is to provide a naming convention
for every metadata considered within the model, in order to be able
to identify one concept in various models or serialisations. The
idea is that by navigating in the model following the logical links
provided, it is possible to construct identifiers called Utypes that
could be understood by any VO tool aware of the model. To avoid
multiplicity, the Utypes are built from the XML schema
representation of the model which already enforces a hierarchical
structure.
 For instance, the size of the sampling element  along the spatial axis in a 2D
image
corresponds to:\\
{\small
Characterisation.spatialAxis.samplingPrecision.samplingPrecisionRefVal.sampleExtent}
The full list of Utypes derived from this model (versions 1.1x of the model) is stored at\\
\violet{\footnotesize{
\url{http://www.ivoa.net/Documents/latest/UtypeListCharacterisationDM.html}}}

\subsubsection{VOTABLE serialisation}
  A VOTABLE serialisation of the characterisation of the IFU MPFS data
set is shown in Appendix C.  Each CharacterisationAxis is shown as a
table, where each property itself is shown as a Group of FIELDS. UML
class attributes are serialised as FIELDS (except if they
have a detailed STC structure; in that case they are translated as a
group of FIELDS). In this example, Utypes are set for each Table,
Group, and Field according to the following rule:
\begin{itemize}
\item[]\textbf{ A Utype is elaborated for each VOTable item in the
serialisation as a string based on instance variable paths in our
object-oriented datamodel.}
\end{itemize}

Other ways of deriving utypes from a valid Xpath to the equivalent
XML element in the XML Characterisation schema have been studied.
The main difference is that this option may use constrained element
(or attribute) values in the Utype path. The IVOA needs to define a
single and robust rule to define this concept. \\


\appendix
\section{ Appendix A: XML serialisation example}
\label{app:xml}
 An XML instance document representing the characterisation of an IFU
data set, taken with the Russian MPFS instrument. It relies on the
XML schema mentioned above. See the corresponding XML document at : \\
\violet{\footnotesize{
\url{http://www.ivoa.net/internal/IVOA/CharacterisationDataModel/MPFS-v1.11.xml}.}}

\section{ Appendix B: VOTable serialisation example}
\label{app:vot} An alternative serialisation, using the VOTable
format and applying the Utype mechanism to map the various items to
the Characterisation Data Model classes and attributes. Utypes are
derived from the Characterisation XML schema as mentioned above. See
the full XML document at : \\
\violet{\footnotesize{\url{http://www.ivoa.net/internal/IVOA/CharacterisationDataModel/MPFSVOt-v1.11.xml}
}}

\section{Appendix C: Characterisation of various dataset properties}
\label{app:tabs}
\begin{table}[bp]
{\small

\begin{tabular}{|l | l | l | l | l | }
\hline
\space \space \textsc{\blue{Axes}}& \textsc{\blue{Spatial}} & \textsc{\blue{Temporal}} & \textsc{\blue{Spectral}}& \textsc{\blue{Observable}}\\
\textsc{\blue{Properties}} &           &                   &                  & \textsc{\blue{e.g. Flux}} \\

\hline \textbf{Coverage}  & & & &  \\ \hline
Location & Central  & Mid-time & Central  & Average flux\\
         &position &           & wavelength &  \\
 & & & & \\
\hline
Bounds  & RA, Dec [min,max] & Start/stop   & Wavelength& Saturation,  \\
         &or Bounding box &    time      & [min, max] & Limiting flux\\
 & [center, size] & & & \\
 & & & & \\
\hline
Support & FOV as  & Time intervals& Wavelength & \\
        & array of polygons & (array) & intervals & \\
 & & & (array)& \\
 & & & & \\
\hline
Sensitivity&  Quantum efficiency &  &Transmission  & Function   \\
            & (x,y)     &       & curve ($\lambda$)     & property \\
 & & & & e.g. linearity \\
 & & & & \\
\hline Filling  & Effective/ &  Live time &  &  \\
  factor        & Total area &  fraction &  & \\
 & & & & \\
\hline
\textbf{Resolution} &  PSF (x,y)  & Duration  & Band  & Flux SNR \\
          &  or its FWHM & per image &FWHM & (stat error)\\
 & & & & \\
\hline
\textbf{Sampling } &  Pixel scale & Duration   & Band       &  (1 ADU \\
\textbf{Precision} &  (x,y)       &per image & FWHM &    equivalent =  \\
& & & &Quantization) \\
& & & & \\
\hline
\end{tabular}
}
    \caption{
\emph{Property versus Axis description of  metadata
    describing a \textbf{2D optical image}. This represents a single
    integration or indivisible stack of exposures, taken in a single
    broad-band filter, so the spectral resolution is the same as the filter
FWHM.  }}
\label{tab:2Dim}
\end{table}


\begin{table}[b]
{\small
\begin{tabular}{|l|l|l|l|l|}
\hline
\space \space \textsc{\blue{Axes}}& \textsc{\blue{Spatial}} & \textsc{\blue{Temporal}} & \textsc{\blue{Spectral}}& \textsc{\blue{Observable}}\\
\textsc{\blue{Properties}} &           &                   &                  & \textsc{\blue{e.g. Flux}} \\

\hline \textbf{Coverage}  & & & &  \\ \hline
Location & Central  & Mid-Time & Central  & Average \\
         &position & & wavelength &  flux \\
 & & & & \\
\hline
Bounds  & Slit   RA, Dec & Start/stop time  &  Wavelength & Saturation,  \\
         & [min, max] or &                  & [min, max] & Limiting \\
 & Bounding box& & & flux\\
  & & & & \\
\hline
Support & Slit as  accurate & Time & Wavelength  & Lowest and \\
        & array of  & (intervals)  &  intervals  & highest  \\
        &polygons   & (array)      &(array)      & value \\
  & & & & \\
\hline
Sensitivity&  Response (x,y) &  & Quantum      & Function   \\
            & along slit    &   &   efficiency & property \\
            &               &   & ($\lambda$)   & e.g.  Linearity\\
 & & & & \\
\hline
Filling  &  Effective/ &  Live time &       &  \\
factor   & Total area & fraction     &       &  \\
 & & & & \\
\hline
\textbf{Resolution} & Slit & Min. extractable & LSF or its & Flux\/SNR \\
          &  area & interval      & FWHM       & (stat error)\\
 & & & & \\
\hline
\textbf{Sampling } &  Slit&  Min. extractable  &Pixel scale  &  (1 ADU \\
\textbf{Precision}  & area & interval  &in $\lambda$    & equivalent\\
& & & &Quantization) \\
 & & & & \\
\hline
\end{tabular}
}

    \caption{\emph{Property versus Axis description of  metadata
    describing a \textbf{1D-Spectrum.}}}
    \label{tab:1Dspec}
\end{table}

\vskip 0.1in

\begin{table}
{\small
\begin{tabular}{|l|l|l|l|l|}
\hline
\space \space \textsc{\blue{Axes}}& \textsc{\blue{Spatial}} & \textsc{\blue{Temporal}} & \textsc{\blue{Spectral}}& \textsc{\blue{Observable}}\\
\textsc{\blue{Properties}} &           &                   &                  & \textsc{\blue{e.g. Flux}} \\

\hline \textbf{Coverage}  & & & &  \\ \hline
Location & Central  & Mid-Time & Central   & Average \\
         &position &          & wavelength &   flux\\
         &         &          & (all spectra) & \\
         & & & & \\
\hline
Bounds  & Field   & Start/stop   & Wavelength & Saturation,  \\
         &  RA, Dec  &   time     & [min,max]  & Limiting \\
 & [min, max]& & (all spectra)& flux\\
         & & & & \\
\hline
Support & Union of fiber  & Time & Disjoint  & Lowest and \\
        & footprints  &  intervals & wavelength  & highest  \\
 & on the sky& (array)& intervals& value\\
          & & & & \\
\hline
Sensitivity&  Response(x,y) &  & Quantum        & Function   \\
            & along         &   &  efficiency  & property \\
            & the slit      &   &($\lambda$)    & e.g.  Linearity\\
  & & & & \\
\hline
Filling  &  Effective/ &  Live time &       &  \\
factor   & Total area &    fraction    &       &  \\
 & & & & \\
\hline
\textbf{Resolution} &  PSF (x,y)  & Min.        & LSF  &  Flux \/ SNR\\
                    &  or its     &  extractable&  or its    & (stat error)\\
                    &   FWHM      &  interval   &    FWHM             & \\
 & & & & \\
\hline
\textbf{Sampling }  &  Pixel scale &  Min.       &Pixel       &  (1 ADU  \\
\textbf{Precision}  & (x,y)        & extractable& scale       & equivalent             \\
                    &              &interval    & in $\lambda$& Quantization)\\
  & & & & \\
\hline
\end{tabular}
}
    \caption{\emph{Property versus Axis description of metadata
    describing \textbf{3D IFU data}. These are taken using a mask of multiple
    slits or  fibres each focusing a separate spectrum onto
    a single detector array. The Support comprises multiple discrete
    intervals in all dimensions, into which data products could be
    decomposed.  The spatial resolution is determined by the telescope
    aperture (and the seeing) which spreads the incident radiation
    over several CCD pixels; the resolution and pixel scales impose
    different constraints on downstream data analysis. }}
    \label{tab:3Dspec}
\end{table}

\begin{table}[htbp]
{\small
\begin{tabular}{|l|l|l|l|l|}
\hline
\space \space \textsc{\blue{Axes}}& \textsc{\blue{Spatial}} & \textsc{\blue{Temporal}} & \textsc{\blue{Spectral}}& \textsc{\blue{Observable}}\\
\textsc{\blue{Properties}} &           &                   &                  & \textsc{\blue{e.g. Flux}} \\

\hline \textbf{Coverage}  & & & &  \\ \hline
Location & Central  & Mid- Time & Central  & Average flux\\
         &position &           & Frequency &  \\
 & & & & \\
\hline
Bounds  & RA,Dec [min,max] & Start/stop   & Frequency  & Saturation, \\
        & or Bounding box & time &[min,max] &rms noise \\
 & [center, size] & & & \\
 & & & & \\
\hline
Support & Primary beam & Time intervals&Frequencies &Peak, \\
        &FWHM & (array) & (array) & 3$\sigma$ rms\\
 & (or mosaic polygons)& & & \\
 & & & & \\
\hline
Sensitivity& Smearing limits/ &Gain- &Bandpass & Dynamic  \\
            & functions (of  integ. &elevation & function(s)    &range  \\
 & time/ chan. width) & &or FWHM(s) & \\
 & & & & \\
\hline Filling  & Fraction &  Live time  & Fraction &  \\
  factor    & of mosaic&  fraction      & above FWHM & \\
 & filled & &sensitivity & \\
& & & & \\
\hline
\textbf{Resolution} & Spatial scales & Min. imageable & FWHM of &RMS noise\\
          &(max and min of  &duration &Hanning  & \\
 &BMaj, BMin, BPA) & &smoothing & \\
& & & & \\
\hline
\textbf{Sampling } &  Pixel scales & Integration  & Channel       &  \\
\textbf{Precision} &  [min, max]       & time &  width &           \\
& & & & \\\hline
\end{tabular}
}

    \caption{\emph{Property versus Axis description of metadata
    describing \textbf{a radio image service}, potentially mosaiced. The
    Max. and Min. spatial resolutions arise from the shortest and
    longest baselines present; any intermediate value may be selected
    when an image is extracted from visibility data.  The spectral
    resolution may be coarsened by smoothing to minimise artefacts.}}
    \label{tab:rad}
\end{table}

\begin{table}
{\small
\begin{tabular}{|l|l|l|l|l|}
\hline
 \textsc{\blue{Axes}}& \textsc{\blue{Spatial}} & \textsc{\blue{Temporal}} & \textsc{\blue{Spectral}}& \textsc{\blue{Observable}}\\
\textsc{\blue{Properties}} &           &                   &                  & \textsc{\blue{e.g. Flux}} \\

\hline \textbf{Coverage}  & & & &  \\ \hline
Location & Central  & Mid- Time & Central  & Average flux\\
         &position &           & Frequency &  \\
 & (0, 0) & (0) & & \\
 & & & & \\
\hline
Bounds  &  Bounding box  & Relative & Frequency  & Saturation, \\
        &  [center, size] & start/stop time &[min,max] &rms noise \\
 && & & \\
\hline
Support & FOV as array & Time interval& Frequencies & \\
        &of polygons &  &  &  \\
 & & & & \\
\hline
Sensitivity&Quantum efficiency&  &Transmission & Detector  \\
            & (x, y) &       & curve  &linearity  \\
 & & & & \\
\hline
Filling  & Effective/ &  (100\%)  &  &  \\
  factor    & Total area&       &  & \\
 &  & & & \\
\hline
\textbf{Resolution} & PSF & Duration & Band &Noise\\
          & FWHM  & &FWHM & error\\
 & & & & \\
\hline
\textbf{Sampling } &  Pixel scales & Duration  & Band       &  Quantization \\
\textbf{Precision} &  [x, y]       &  &  FWHM &           \\
\hline
\end{tabular}
}
    \caption{\emph{Property versus Axis description of metadata
    describing a \textbf{simulated CCD observation} in a single band. The
    spatial coordinates may be expressed in (x, y) independent of
    celestial position.}}
    \label{tab:sim}
\end{table}

\vskip 0.1in

\clearpage

\section{Appendix D: Requirements for Data Model compliance}
\subsection{Limitations in this version}\label{sec:Mlim}
The first three levels of Characterisation are now fully described
and take explicit values. The fourth level of the structure can
contain functions (e.g. the variation of noise with position) or
URLs (e.g. the location of a weight map). Data providers may have
varying expectations about how these advanced metadata should be
delivered, so we will expand the description of this level in a
future version of the model, after polling the community for the use
of weight maps, variability maps, etc... We anticipate that the
first three levels will answer more than 70\% of present needs.

It is not yet possible to implement rules linking coverage on
different axes. For example, if a survey consists of spatially
distinct fields, observed in several wavebands, but there are fields
which do not contain all wavebands, then each field and/or each
waveband must be described separately. Similarly, separate
descriptions are required if resolution or noise (for instance)
behave differently in various areas of Support.

\subsection{Implementing Characterisation}
\subsubsection{Data Providers}
Several tools are being developed to assist data providers supply
 metadata.  These include extraction of information from FITS headers and
 a form interface called CAMEA which allows the user to enter values for
 Characterisation elements and translates this to XML. We will also provide
 XML templates for manual editing. We will investigate what would be
more convenient for large data collections depending on how they
store their existing metadata.

Metadata required by Characterisation might be extracted from a
number of sources such as:
\begin{itemize}
    \item An archive database;
    \item An observing log
or other description which might be stored in a database or as
ascii, xml or other documents;
    \item FITS headers, which provide
more or less direct routes:
\begin{itemize}
    \item Unambiguous identification between e.g. a database column or
FITS keyword and a Characterisation element;
    \item Correspondence
with formulaic modification, e.g. adding explicit units or
calculating the field of view of an interferometer;
    \item A separate information source e.g. resolution of the telescope using different
frequencies/configurations;
    \item Offine/human memory/judgement
\end{itemize}
\end{itemize}

The following sections outline our proposals for which of the status
strings MANDATORY, RECOMMENDED or OPTIONAL should be applied to each
element of the XML schema. The status strings are used as in the
SIAP proposed recommendation
(\violet{\footnotesize{\url{http://www.ivoa.net/Documents/WD/SIA/sia-20040524.html}}}),
interpreted as follows:
\begin{itemize}
    \item MANDATORY means that the metadatum is fully required to make
    the data usable in
    basic VO services
    \item RECOMMENDED means that this item should be given if at all
    possible to improve the interpretation of the data or their use in
    a wider range of
    VO services
    \item OPTIONAL means that such metadata elements would help to
    give more precise interpretation but are not vital.
\end{itemize}

An implementation may have one among the following level of
compliance and be :
    \begin{itemize}
        \item "partially compliant", if it implements some (but not all)
        MANDATORY/MUST elements
        \item "compliant" , if it implements all MANDATORY/MUST
        elements
        \item "fully compliant", if it implements \textbf{all} MANDATORY/MUST
        elements and \textbf{all} RECOMMENDED/SHOULD ones
    \end{itemize}
The prime goal is to get this model applied by data providers in
useful ways. We should make it as easy as possible to describe any
kind of observed or simulated data by minimising the number of
compulsory fields. At the same time we must encourage data providers
to give enough information to expand the ways in which data can be
selected or manipulated by VO tools currently or imminently
available.

\subsection{Requirements for compliance}
\subsubsection{General considerations}
 On each axis, the first three levels (Location, Bounds, Support), must
be given explicit numerical values (or arrays of values) in order to
be accessible to any tool. Other elements may be given numerical
values, or functions, or indirect references (URIs) but these are in
general not used at present.

Users are strongly encouraged to evaluate coarser levels of
description explicitly even if they also provide finer levels. We
need to decide what users do if they are not giving a value for an
element e.g. leave blank, consistent with other models.

Location, Bounds and other higher Characterisation levels are
intentionally approximations to provide a simple inclusive
description of the data.

The Location value may be determined with some error, which might be
mentioned inside the STC structure used for Coordinates. This is to
be distinguished from Bounds which should be the outer limits to
anywhere data might be found.

Accuracy properties describe uncertainties in the mapping process of
data values along axes, see Section ~\ref{sec:Macc}.

The values for some elements must be given as arrays, defined as in
STC, and the required number of arguments must be present if any
are. Bounds, for instance, describes a unique region on an axis as
e.g. $(\alpha1, \delta1; \alpha2, \delta2)$, whilst the
ResolutionSupport is given relatively e.g. telescope beam major and
minor axes in arcsec and position angle.

\subsubsection{Defaults}\label{sec:defaults}
Defaults might sometimes be possible for values which have not been
provided. We do not think that such defaults should be coded into
the description, rather that software which looks for the value of a
missing element might be able to make an intelligent assumption. It
is up to the writers of a software tool specification to decide
whether it is more dangerous to use defaults and risk a lower level
of accuracy, or to ignore data which is not adequately specified and
thus loose potentially important information.

  For example, if \textbf{Location}
is not given then, for some axes, software may take the default
\textbf{Location} as the mid-point of the \textbf{Bounds}
\footnote{this might be complicated (e.g. some spatial coordinates)
or impossible }.

If \textbf{Bounds} are not given then e.g. if a spatial axis has
Coordsys ICRF some software might assume all-sky coverage
\footnote{a more restricted coverage might be derived once there is
a link to Observation and the telescope location}.


If \textbf{Support} is not given \textbf{Bounds}, if present, could
be used.

If the \textbf{unit} or \textbf{Coordsys} element is not given for
any level, the values for the \textbf{CharacterisationAxis} are
used; be careful, as this may be unsuitable (e.g. if the
\textbf{CharacterisationAxis} units are sexagesimal, then a single
number for an error could be in degrees or arcsec or ...).

\subsection{Axes}
 It is MANDATORY to provide at least one axis (coded as a
\textbf{CharacterisationAxis} element).  All three of the Space,
Time and Spectral Coverage axes are RECOMMENDED\footnote{some might
be considered irrelevant for simulated data, or not conventionally
provided e.g. for old spectra with no time stamp}.

The unit and coordinate system are MANDATORY for each Axis present.
These may be relative to an internal reference only, e.g. pixel
spatial coordinates. In such a case both the \textbf{Location} and
\textbf{Bounds} are MANDATORY for that axis. Note that STC allows
`RELOCATABLE', for example as a valid \textbf{Location} for
simulated data, unless this is incompatible with the specified
coordinate system.

Space-, Spectral- and Time-related axes, and most other potential
axes, are already defined in STC; where this is the case, it is
MANDATORY to use the STC coordinate system and unit
definitions.
Various cases of how to re-use STC elements are shown
in the example XML documents provided.
The
\textbf{\emph{Observable}} Axis is RECOMMENDED \footnote{its
omission may seem reasonable for e.g. the coverage intended for a
future survey}.

Axes which are not yet defined in STC (such as Polarization at the
present time) are OPTIONAL but a reference to the definition of the
proper Coordinate System should be given.

\subsubsection{Axis Flags}
  For each \textbf{CharacterisationAxis} element (spatial, spectral, observable, etc.):\\
A flag to indicate if it is an independent or a dependent variable
(`true' or
`false') is RECOMMENDED.\\
 A flag to indicate its calibration status is RECOMMENDED:\\
 CALIBRATED, UNCALIBRATED, RELATIVE, NORMALIZED; default UNCALIBRATED.\\
Flags to indicate SamplingStatus are OPTIONAL (these are RECOMMENDED
    where they are customarily relevant):
\begin{itemize}
    \item undersamplingStatus (`true' or `false')
    \item regularsamplingStatus (`true' or `false')
\end{itemize}

\vspace{1.5cm}

\subsection{Coverage}\label{sec:Mcov}
For each CharacterisationAxis, it is MANDATORY to give either the
\textbf{Location} or the \textbf{Bounds} elements. Both
\textbf{Location} and \textbf{Bounds} are
RECOMMENDED if these are available.\\
\textbf{Support} is RECOMMENDED; if it is given then it is MANDATORY
also to give the \textbf{Bounds}\footnote{If different areas of
Support apply on different axes, a separate description should be
used at the level where each subset of data can be described
unambiguously,
see Section ~\ref{sec:Mlim}}.\\
 \textbf{Sensitivity}\footnote{Here, Sensitivity is the dependence of
a detector response or equivalent with position on the given axis.
This is not the limiting sensitivity in the sense of the faintest
detectable flux, which is given by the lower Bound of the Observable
axis.} (e.g. the URI of a weight map, or a function) is
OPTIONAL.\\
 The Unit and/or CoordSystem is OPTIONAL  for each of these
coverage layers; if not given, they will default to the units and
CoordSystem used for the \textbf{CharacterisationAxis} element (i.e.
when the axis was first defined).

\subsection{Other Properties: Resolution and Sampling Precision}
\label{sec:MresolSampl}
\textbf{Resolution} and \textbf{SamplingPrecision} relate to a
specific \textbf{Coverage} along one CharacterisationAxis. They are
organised according to progressive levels of description as in
\textbf{Coverage} but themselves contain the relevant layers, e.g.,
for some axis:
 \begin{itemize}
    \item at level 1: \textbf{resolutionRefVal} instead of Location stands for a typical or
     average value for the resolution as in Spectral.Resolution.resolutionRefVal
    \item at level 2: \textbf{Bounds} contains the lowest and highest values present
    as    in Spatial.Resolution.resolutionBounds
    \item at level 3: \textbf{Support} represents sets of discrete ranges of sampling    intervals
    as in Spectral.SamplingPrecision.Support
    \item at level 4: \textbf{resolutionVariability} stores the variability of resolution with
    position
    on the axis as in Spatial.Resolution.Variability
\end{itemize}

If there are many areas of \textbf{Support} within the
\textbf{Coverage}, the \textbf{Accuracy}, \textbf{Resolution} and
\textbf{SamplingPrecision} should refer to the inside of
each Support area. However, in this version of the model, it is
assumed that, in principle, on any one axis, one description of each of these
properties applies to all Support areas, otherwise each area must be
described in a separate Characterisation tree description (see
Section ~\ref{sec:Mlim}).

The \textbf{Resolution} and/or \textbf{SamplingPrecision} are
OPTIONAL; if they are present, it is MANDATORY to give the unit and
Coordsys on axes where the units of the CharacterisationAxis would
not make sense or are ambiguous; otherwise the CharacterisationAxis
values are used. The unit and Coordsys are OPTIONAL for any level of
\textbf{Accuracy}, \textbf{Resolution} or \textbf{Sampling},
otherwise the value defined at the start of the \textbf{Accuracy},
\textbf{Resolution} or \textbf{Sampling} definitions is used.

\subsubsection{Resolution}\label{sec:MResol}
If \textbf{Resolution} is present, then it is MANDATORY to give the
\textbf{ResolutionRefVal} (i.e. Location). \textbf{ResolutionBounds}
are  RECOMMENDED. The \textbf{ResolutionSupport}. and
\textbf{ResolutionVariability} (as a function of position on that
axis) are OPTIONAL.

\subsubsection{Sampling Precision}\label{sec:MSampl}
If \textbf{SamplingPrecision} is present, it is MANDATORY to give a
\textbf{samplingPrecisionRefVal} (i.e. Location) which contains both
\textbf{samplingPeriod} and \textbf{sampleExtent}. It is MANDATORY
to provide the \textbf{samplingPeriod}, whilst an explicit
\textbf{sampleExtent} is RECOMMENDED  but it is not required.

\textbf{SamplingPrecisionBounds}, \textbf{SamplingPeriodLimits} and
 \textbf{sampleExtentLimits} are also RECOMMENDED.

The \textbf{SamplingPrecision.Support} and related values for the
 samplingPeriod and/or the sampleExtent are OPTIONAL.  The
 \textbf{samplingPrecision.Variability} (i.e. Sensitivity) (in the
 form of a samplingPrecisionMap to describe variations along an axis)
 is OPTIONAL.

The \textbf{FillFactor} is RECOMMENDED for any axis where the actual
coverage in each Support region is significantly less than 1 but the
filling is too complex to be described practically using
Sampling\footnote{FillFactor applies to the usable fraction of data
within each Support area, as presently defined. If we find that the
majority of users want it to be the useful fraction of the whole
Bounds, the name and definition will be changed in a future
version.}.

 The \textbf{FillFactor} of the \textbf{SamplingPrecison} is OPTIONAL;
 if it is present and if \textbf{SamplingPeriod} and
 \textbf{SampleExtent} are also given, then logically:\\
$FillFactor = SampleExtent/SamplingPeriod$ \\and the data provider
should take care that the values and units given are consistent with
this relationship.

\subsection{Accuracy}\label{sec:Macc}
\textbf{Accuracy} values for the precision of measurements are
RECOMMENDED for each CharacterisationAxis, divided into statistical
and systematic uncertainties (or appropriate alternative definitions
of uncertainties). For each CharacterisationAxis where
\textbf{Accuracy} is provided:

\begin{itemize}
  \item It is MANDATORY to give the unit and Coordsys on axes where
the units of the CharacterisationAxis would not make sense or are
ambiguous, otherwise the CharacterisationAxis values are used.
\item The unit
and Coordsys are OPTIONAL for any axis\footnote{for example
normalised units such as a flux accuracy of 0.03 given flux
measurement.}.
\item It is
MANDATORY to give the \textbf{ErrorRefVal} (typical value).
\item The \textbf{ErrorBounds} are OPTIONAL for uncertainties which vary along
the domain of the axis.  \item The URI of an \textbf{ErrorMap} which
describes the variation of errors with location is OPTIONAL.
\end{itemize}

\section{Appendix E: Updates of this document}
%
%

\begin{itemize}
\item version 0.9
	\begin{itemize}
    \item full re-organisation, rewriting and simplification of the Model description, Anita
Richards April/May 2006
		\item new XML schema, F.Bonnarel
	\end{itemize}
\item version 0.93
	\begin{itemize}
    \item update XML schema and use STC blocs elements
  	\item add examples of characterisation metadata for various data
sets and various dimensions
	\end{itemize}
\item version 0.96 to 0.99
\begin{itemize}
    \item finalise the XML schema and interface it with the STC XML schema
  	\item update the document format according to the IVOA master document.
\end{itemize}
\item version 1.0
	\begin{itemize}
    \item revision by the authors. Fill factor, Support discussions: minor formulation changes
    \item add Appendix D: Requirements for Data Model compliance
prepared by Anita Richards and discussed at the Moscow interoperability
meeting
	\end{itemize}
\item version 1.1
        \begin{itemize}
            \item modify data model for more compatibility to Spectrum data model :
        remove AxisFrame class not present in Spectrum DM.
	        \item modify schema and figures accordingly
        \end{itemize}
\item version 1.11
        \begin{itemize}
            \item make the document compatible to IVOA referencing mecanism:
            update links.
            \item list up changes of the various draft versions in the present document
        \end{itemize}
\item version 1.12\\
    Changes following the RFC and tcg comments and requirements
        \begin{itemize}
            \item Schema reference:update links to XML schema 
\\ \violet{\footnotesize{\url{http://www.ivoa.net/xml/Characterisation-v1.11}}}
            \item update ref to examples MPFS : \\ 
\violet{\scriptsize{\url{http://alinda.u-strasbg.fr/Model/Characterisation/examples/MPFS-v1.11.xml}}}
            \item  change Fig.1 about interactions between Characterisation and 
other models.
        Observation class gets more obvious and inclusion of Char inside Observation DM too.
            \item update reference to the Quantity DM: simplify
            \item introduce Spectrum DM reference in section \ref{sec:othermodels} 'Links to other efforts'
            \item include a Scope section at the beginning of the document
            \item Appendix E: include the document revision history directly in this file instead of pointing to an external file
            \item use smaller fonts for figure captions 
        \end{itemize}
        March 2008, 25: corrected the links to the UtypeList document
\end{itemize}

\end{document}